\documentclass[fleqn,usenatbib]{mnras}
\usepackage{fix-cm}
\usepackage{tabularx}
\usepackage{hyperref}

\usepackage[T1]{fontenc}
\usepackage{pdflscape}
\usepackage{tikz}
\usetikzlibrary{tikzmark}

\DeclareRobustCommand{\VAN}[3]{#2}
\let\VANthebibliography\thebibliography
\def\thebibliography{\DeclareRobustCommand{\VAN}[3]{##3}\VANthebibliography}

\usepackage{graphicx}	
\usepackage{amsmath}	
\usepackage{amssymb}	

\title[Emergent Denoising of SDSS Galaxy Spectra]
      {Emergent Denoising of SDSS Galaxy Spectra Through Unsupervised Deep Learning}

\author[Camilleri et al.]
{Oliver~C. Camilleri$^{1}$\thanks{E-mail: oc00149@surrey.ac.uk}, Zahra Sharbaf$^{2,3}$, Ignacio Ferreras$^{4,2,3}$\thanks{Corresponding author: i.ferreras@ucl.ac.uk}\\
$^1$ Centre for Vision, Speech and Signal Processing (CVSSP), University of Surrey, Guildford, Surrey, GU2 7XH, UK\\
$^2$ Instituto de Astrof{\'i}sica de Canarias, Calle V{\'i}a L{\'a}ctea s/n,
E38205, La Laguna, Tenerife, Spain\\
$^3$ Departamento de Astrof{\'i}sica, Universidad de La Laguna, E38206 La Laguna, Tenerife, Spain\\
$^4$ Department of Physics and Astronomy, University College London, London WC1E 6BT, UK
}

\date{Revised version (Rv2), \today}

\pubyear{2025}

\begin{document}
\label{firstpage}
\pagerange{\pageref{firstpage}--\pageref{lastpage}}
\maketitle

\begin{abstract}
Spectroscopy represents the ideal observational method to maximally extract information from galaxies regarding their star formation and chemical enrichment histories. However, absorption spectra of galaxies prove rather challenging at high redshift or in low mass galaxies, due to the need to spread the photons into a relatively large set of spectral bins. For this reason, the data from many state-of-the-art spectroscopic surveys suffer from low signal-to-noise (S/N) ratios, and prevent accurate estimates of the stellar population parameters. In this paper, we tackle the issue of denoising an ensemble by the use of unsupervised Deep Learning techniques trained on a homogeneous sample of spectra over a wide range of S/N. These methods reconstruct spectra at a higher S/N and allow us to investigate the potential for Deep Learning to faithfully reproduce spectra from incomplete data. Our methodology is tested on three key absorption line strengths and is compared with (noiseless) fitted data to assess retrieval biases. The results suggest a standard Autoencoder as a very powerful method that does not introduce systematics in the reconstruction. We also emphasise the need for careful analysis, demonstrating that classical signal-processing methods like Butterworth filters can yield spectra that appear smoothed yet deviate significantly from the true, underlying signals. Denoising methods with minimal bias will maximise the scientific return of ongoing and future spectral surveys such as DESI, WEAVE, or WAVES. 
\end{abstract}



\section{Introduction}
\label{Sec:Intro}

Galaxy spectra in the wavelength interval from near ultraviolet to near infrared encode a vast amount of information regarding the properties of the underlying stellar populations. The continuum and absorption lines of the photospheres of constituent stars leave their imprint on the integrated spectra, and represent the workhorse of galaxy formation studies concerning the star formation and chemical enrichment histories. Spectroscopic surveys of galaxies, such as the Sloan Digital Sky Survey \citep[SDSS,][]{York:00} have greatly helped deepen our understanding of galaxy formation. However, spectroscopy requires large integration times, as the faint light from galaxies is spread with respect to wavelength. Often times, spectra have been used -- mainly by cosmologists -- as a tool to derive redshift and thus determine the large-scale distribution of galaxies. For instance, targets of signal-to-noise S/N$\sim$5 in telltale emission lines are good enough to secure a redshift, especially if doublets can be discerned, such as [O{\sc II}] at $\lambda\sim 3727$\AA.
However, exploring the galaxies themselves through their stellar populations requires deeper data, at a higher S/N ratio, in order to compare the observations with detailed models of stellar population synthesis \citep[see, e.g.][]{Walcher:11,Conroy:13}. In this regard, it is not uncommon to need values of S/N per resolution element above $\gtrsim$20-30, if not higher, for detailed analyses of subtle differences in the populations, such as variations of chemical abundances or the initial mass function \citep[e.g.][]{FLB:13,FLB:17,IF:19,Woo:24}. In this regard, spectroscopic surveys tend to be optimised to produce large volumes of data, at the cost of a lower S/N, and so, any algorithm aimed at increasing the S/N of the data proves a very valuable tool, especially with the ongoing and upcoming surveys such as DESI \citep{DESIDR1}, WEAVE \citep{Weave} or WAVES \citep{Waves}.

The actual S/N of an observation is borne out of the competition between the photons from the galaxy and spurious photons or counts coming from unwanted sources, such as the background sky, airglow, detectors and reduction artifacts, etc. Increasing the S/N of a single spectrum is typically not feasible unless models are used, therefore introducing substantial -- and sometimes unknown --  systematics. Our approach to this problem starts from a large ensemble of spectra taken by the same instrument and reduced by the same data pipeline. The ensemble should also include a large amount of high quality (i.e. high S/N) spectra, so that the method can somehow interpolate among the ensemble members to produce optimised versions of the data. Traditional data-driven methods, such as Principal Component Analysis (PCA), have been applied to stellar and galaxy spectra, for instance to remove the emission lines from airglow \citep{WH:05}. These methods are commonly used for classification purposes \citep{Madgwick:03,McGurk:10} and can also help in the interpretability of the information content, for instance by  exploring the resultant latent space \citep[e.g.,][]{Rogers:10a,ZS:25}. However, as a linear method, PCA is less versatile to encode and model the many intricacies of the spectra in a large ensemble.

Deep Learning (DL) methods are seeing a rapid uptake in use throughout astronomy. As powerful universal function approximators \citep[][]{Hornik}, DL models may address problems that traditional data-driven approaches struggle with. Indeed, there are now a multitude of works which demonstrate their broad applicability within galaxy and stellar spectra analysis. Applications include classification \citep[e.g.,][]{Folkes:96,CS:99,Wu:24}, dimensionality reduction \citep[e.g.,][]{Portillo:20}, recovery of spectra with bad quality \citep[e.g.,][]{Wang:17} and the search for anomalies \citep[e.g.,][]{Baron:17,Liang:23}. In this paper, we adopt a set of DL algorithms, each trained on a large set of galaxy spectra from the Legacy part of SDSS, and then assess their ability to increase the S/N of input spectra. These models are unsupervised, and in contrast to works like \citet{Scourfield:23}, do not rely on adding synthetic noise to \textit{training} data. Instead, the denoising effects seen emerge directly from the choice of objective function and the statistics of real spectra which are subject to a range of information bottlenecks. Furthermore, we experiment with the reconstruction of spectra from incomplete data and attempt to explain predictions by examining how models leverage spectral features to make decisions – an important endeavor given the increasing adoption of black box models.

A key requirement for any denoising approach is that it must not introduce systematic biases into the sample. In particular, increasing the S/N ratio should not come at the cost of altering the underlying spectral information. Unlike traditional smoothing techniques, which typically improve S/N by degrading spectral resolution, our approach seeks to preserve the intrinsic structure of the spectra while reducing noise. Several studies have begun to explore these challenges using DL frameworks.  For example, \citet{Lovell:19} designed their training procedure such that the model effectively learns to look past noise. Synthetic spectra are made more realistic by injecting artificial noise at a given S/N ratio, and multiple noise realisations are generated for each spectrum. This resampling strategy prevents the model from overfitting specific noise patterns and significantly improves predictive accuracy, allowing the network to recover star formation histories despite noisy inputs. While the primary goal of \citet{Portillo:20} is dimensionality reduction using a variational autoencoder (VAE), several aspects of their methodology make the model inherently noise-aware. The reconstruction loss explicitly incorporates per-pixel observational uncertainties, effectively down-weighting noisy regions of the spectrum, while an imposed uncertainty floor prevents high S/N pixels from dominating the training. In addition, problematic regions such as bad pixels are preprocessed via PCA-based infilling to avoid propagating artifacts into the learned representation. Similarly, \citet{Melchior:23}, enforce robustness by weighting their reconstruction loss using inverse variance from the SDSS data pipeline. Their generative {\sc SPENDER} model is designed to capture the full diversity of galaxy spectra across different redshifts and instrumental resolutions. {\sc SPENDER} encodes spectra into a physically interpretable latent space and also treats redshift as a free parameter, as the input data is fed in the observer frame. Work on spectral denoising has been presented in \citet{Scourfield:23}, mainly focused on retrieval of emission line data. The authors conclude that a variational autoencoder (VAE) performs better than PCA, and study the effect of denoising
DESI data from an SDSS-trained set regarding the relation-ship between stellar mass and (gas phase) metallicity. This paper focuses on a quantification of the performance of related ML-methods in reducing the noise in the absorption features, an essential observable when constraining stellar population properties, and assess the potential overfitting-related bias in the reconstruction, by comparing these ML algorithms with a classical filtering method that works on single spectra and therefore does not use the inherent information from the ensemble.

The structure of the paper is as follows, we give a brief presentation of the working sample in \S~\ref{Sec:Sample}, along with a description of the various methods tested for denoising in \S~\ref{Sec:Method}. The comparison of retrieved and original data is shown in \S~\ref{Sec:Comp}, with a statistic that quantifies the residual with respect to fitted data or with respect to the original noisy spectra. A brief discussion of the explainability of the DL performance is presented in \S~\ref{Sec:explainability}. Finally, our concluding remarks are given in \S~\ref{Sec:Conclusions}.

\section{Preparing the sample}
\label{Sec:Sample}

We select our working sample from the Legacy set of galaxy spectra of the Sloan Digital Sky Survey \citep{York:00}. The data include a large number of spectra (of order 1 million), and most importantly covers a wide range of S/N, with a large number of high quality data at a S/N higher than 10-20\,\AA$^{-1}$ (averaged in the SDSS-$r$ band), and many spectra at lower S/N. This represents an ideal training sample as the data processing is homogeneously performed, minimising biases, and covering all types of evolutionary stages, mass, morphology, etc. Moreover, each observation includes, in addition to the actual spectrum, a best fit model that can be adopted as a synthetic case to which noise is added, as we will show below. The sample and its preparation for the analysis is identical to the one presented in \citet{ZS:23} as well as the motivation behind the constraints regarding the quality of the data. We refer the reader to that paper for more details on sample preparation, but the essential selection is presented here. The sample of spectra is taken from SDSS Data Release 16 \citep{SDSSDR16}, and corresponds to single fibre measurements of the central parts of galaxies (3\,arcsec diameter), at spectral resolution ${\cal R}\sim 2,000$ \citep{Smee:13}. The targets are selected as completely as possible down to a Petrosian flux level for the target galaxy in the SDSS-$r$ band of $r<17.77$\,AB \citep{Strauss:02}. In our working sample a constraint is imposed in stellar velocity dispersion between 100 and 150\,km\,s$^{-1}$ as a wider range will also introduce the expected bias in effective spectral resolution caused by the kinematic kernel. Denoising methods will eventually take into account more general samples, but this restriction allows us to assess the simplest scenario. The data are further constrained in redshift, z$\in$[0.05,0.1] and in S/N measured as an average within the SDSS-$r$ band higher than 10 per pixel ($\Delta\log(\lambda/$\AA$)=10^{-4}$). The total sample comprises 68,794 spectra, which were retrieved from the main SDSS database, de-redshifted and de-reddened. The de-redshifting process simply makes use of the official SDSS estimate of the redshift ($z$), performing a flux-conserving interpolation from observed wavelength ($\lambda_O$) to rest-frame wavelength ($\lambda_R$) following the standard relation: $\lambda_R=\lambda_O/(1+z)$. The de-reddening also follows standard procedure, with an (oberved) wavelength-dependent multiplicative factor adopting the standard \citet{Fitz:99} attenuation law. We note this correction only concerns the foreground (Milky Way) reddening, leaving the contribution of the dust intrinsic to each galaxy.

To remove the variations caused by different stellar mass and redshift of the galaxies, all spectra are normalized to the same average flux in 6,000-6,500\AA\ window, in the rest-frame ($\lambda_R$). This ensures that the flux variations among spectra only relate to the features in them, and not to the expected changes caused by distance or total stellar mass.  The spectral range is restricted to the rest-frame wavelength $\lambda_R\in[3700,7500]$\AA, that features a large number of absorption and emission features, and -- from a Shannon-entropy point of view -- keeps most of the
information content regarding the underlying stellar populations \citep{Entropy}. Finally, for one of the tests, we explore the denoising procedure when training on data without continuum. For that purpose, we take the robust high percentile method of \citet{Rogers:10b} to define the continuum that is removed from each spectra for this test case. We refer the reader to this paper for details, but in a nutshell, this method defines the continuum in each spectral bin as the 90th percentile of the flux distribution within a relatively small spectral window (100\AA), centered in each bin. For more details about the sample, please see \citet{ZS:23}.

\begin{figure}
\includegraphics[width=85mm]{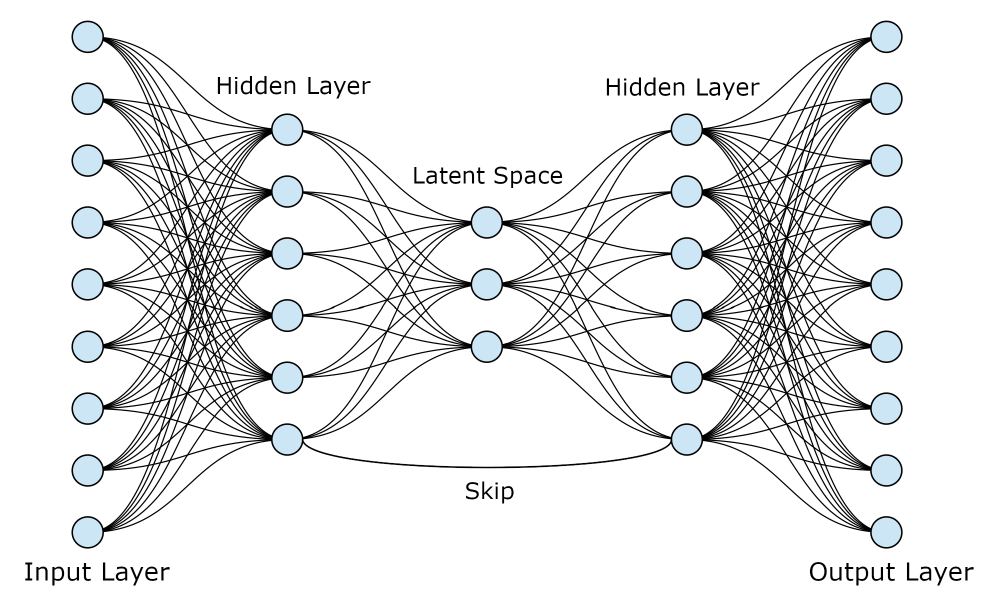}
\caption{A 9-6-3-6-9 autoencoding network with an added skip connection between two hidden layer neurons. The input is embedded within a learned space of lower-dimensionality known as a latent space. From this representation, the input is reconstructed in the output layer.}
\label{fig:network}
\end{figure}
\section{Denoising the spectra}
\label{Sec:Method}

In order to devise a denoising methodology, one can follow two approaches: a ``classical'' one, which operates on individual spectra (i.e., without ensemble learning), or a standard machine learning technique that leverages the ensemble properties to learn the intrinsic noise model of the sample. Classical methods based on resampling or smoothing can, in principle, increase the S/N, but they do so by crudely suppressing high-frequency information. Our goal is to preserve as much physically relevant structure in the denoised data as possible. 
In \S\ref{Ssec:DL}, we adopt a set of ensemble learning methodologies that denoise spectra in an unsupervised manner. In addition, as an illustrative example of a classical procedure, we present in \S\ref{Ssec:BF} the Butterworth filter, which provides a smooth, tunable frequency cutoff while minimising distortion. Code for training all of our deep models can be found on \href{https://github.com/pipyat/DenoisingGalaxySpectra}{GitHub}.

\subsection{Deep Learning}\label{Ssec:DL}
As truly random noise is inherently incompressible, we developed four different deep models, each tasked with reconstructing spectra through some form of information restriction. We emphasize here that training only uses the original SDSS spectra, not the best fits (with or without noise), explored further below. The full spectrum (FS) model reproduces spectra using the entire spectrum as input. In contrast, the narrow windows (NW) and continuum subtracted (CS) models perform this reconstruction using just two specific wavelength windows and continuum-removed spectra, respectively. The windows in question, defined in \citet{ZS:23}, span 3800-4200\AA\ and 5000-5400\AA. Going forward, the former will be referred to as the ''blue'' region while the latter will be referred to as the ''red'' region. FS and CS possess autoencoder-like bottleneck architectures \citep{TZ:15}. Autoencoders are a type of unsupervised feed-forward neural network that reduce the dimensionality of input data by learning to embed it within a compressed, abstract space known as a latent space. This kind of data compression network is illustrated in Fig.~\ref{fig:network}, while our exact network configurations are provided in  Tab.~\ref{Tab:models} . 

\begin{table}
\begin{center}
\caption{Network architectures for our full spectrum (FS), narrow windows (NW), narrow windows-skip (NW-S), and continuum subtracted (CS) models. $\oplus$ denotes layer concatenation and inputs/outputs are in bold. The red arrows indicate a skip connections between layers.}
\begin{tabular}{ |c|c|c|c| } 
\hline
\textbf{FS} & \textbf{NW} & \textbf{NW-S} & \textbf{CS} \\
\hline
\textbf{3800} & \textbf{400  400} & \textbf{400  400} & \textbf{3800} \\
3000 & 400 400 & \tikzmarknode{startL}{400}  \tikzmarknode{startR}{400} & 2800 \\
2100 & 800  800 & 800  800 & 1200 \\
1500 & 1100 1100 & 1100 1100 & 800  \\
850  & $\oplus$ & $\oplus$ & 450 \\
350  & 1800 & 1800 & 400 \\
110  & 2100 & 2100 & 450 \\
300  & 2400 & 2400 & 800 \\
750  & 2800 & \tikzmarknode{endR} {\tikzmarknode{endL}{2800}} & 1100 \\
1400 & 3000 & 3000 & 1600 \\
2700 & 3400 & 3400 & 2800 \\
\textbf{3800} & \textbf{3800} & \textbf{3800} & \textbf{3800} \\
\hline
\end{tabular}
\label{Tab:models}
\end{center}

\begin{tikzpicture}[overlay,remember picture]
  \draw[->,thick,red]
    (startL.west) -- ++(-0.6em,0) |- (endL.west);

  \draw[->,thick,red]
    (startR.east) -- ++(0.6em,0) |- (endR.east);
\end{tikzpicture}
\end{table}

We probe the influence of skip connections using the narrow windows-skip (NW-S) model. Skip connections, illustrated in Fig.~\ref{fig:network}, are shortcuts that allow unmodified information to flow from earlier to later layers, ensuring crucial information is not lost and potentially helping networks learn more useful representations. A theoretical exploration of skip connections and their benefits can be found in \citet{skip}.

\begin{figure}
\includegraphics[width=85mm]{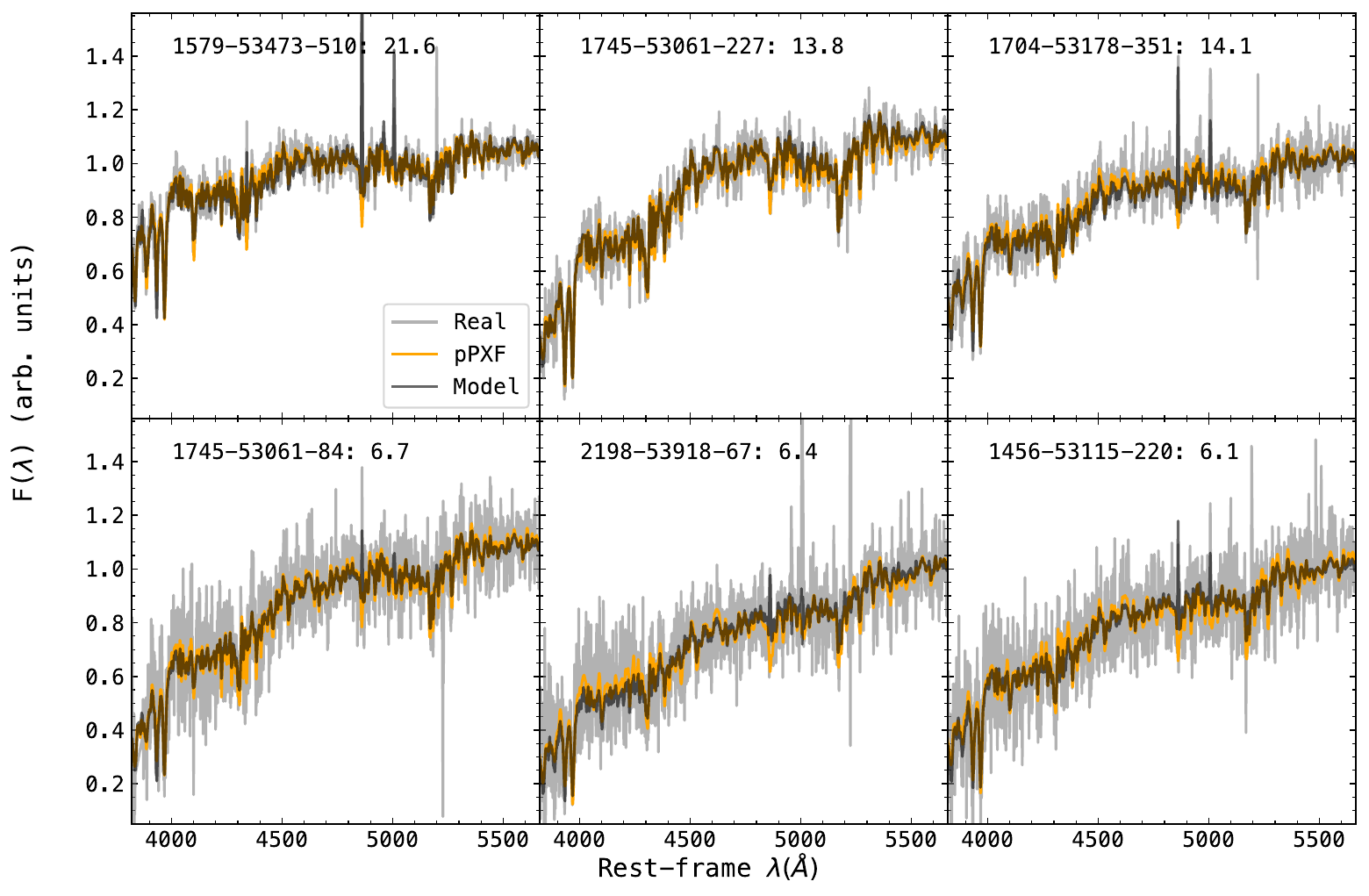}
\caption{Illustration of the best fit models from the official SDSS data (black lines), with respect to the actual data (grey lines). We also show, as reference (orange lines), a standard spectral fitting model in each case, using {\sc pPXF} \citep{ppxf}. See text for details.}
\label{fig:SDSSFits}
\end{figure}

We use either ReLU or PReLU activation functions for all layers other than those corresponding to outputs, which are linear. The ReLU function is defined as
\begin{equation}
\text{ReLU}(x) =
\begin{cases}
0, & x < 0 \\
x, & x \geq 0
\end{cases}
\end{equation}
PReLU was used only in the CS case, as unlike ReLU, it can meaningfully represent negative values. It is defined as
\begin{equation}
\text{PReLU}(x) =
\begin{cases}
\alpha x, & x < 0 \\
x, & x \geq 0
\end{cases}
\end{equation}
where $\alpha$ is a learnable parameter.
In order to minimise prior assumptions about errors or noise, we adhere to a standard reconstruction loss; the objective function of choice for all three models was a mean absolute error
\begin{equation}
    MAE = \frac{1}{N}\sum_{i=1}^{N}\left|\hat{S_{i}}-S_{i}\right|
    \end{equation}
where $\hat{S}_{1...N}$ are predicted spectra and $S_{1...N}$ are the originals. Crucially, the MAE reconstruction objective is robust to training sets containing outliers and noisy data points because it penalises all errors linearly. In contrast, the more commonly used mean squared error (MSE) treats the differences between reconstructions and targets quadratically, allowing outliers to dominate the training signal. We demonstrate the advantage of adopting MAE over MSE in this specific context within Appendix~\ref{app:appendixA}. While promising results have been produced using MSE in previous works, such as \cite{Portillo:20}, our work explores a slightly different reconstruction problem; as the objective is to denoise, we do not want the output to exactly match the input. A theoretical study on MAE's benefits in vector-to-vector regression settings can be found in \citet{mae_benefit}, which also experimentally illustrates its behaviour in noisy settings.
\\
We optimise all models using the Adam method \citep[as defined in][]{adam} and an initial learning rate of $10^{-4}$. This is reduced during training via adaptive learning rate decay with a reduction factor of 0.78. Training was stopped once the validation loss plateaued, decreasing by less than $10^{-6}$ over 5 consecutive epochs.

\subsection{Butterworth Filtering}\label{Ssec:BF}
The Butterworth filter (BF) is a classical signal processing approach used to selectively attenuate unwanted frequencies within a general signal. Its maximally flat frequency response in the passband suppresses signal distortion, making it a popular choice for denoising in a range of domains. The squared magnitude of the frequency response, $|H(\omega)|^2$, at angular frequency\footnote{In this context, the angular frequency is the Fourier-equivalent dual variable to wavelength, therefore $\omega\propto\lambda^{-1}$.}, $\omega$, is given by
\begin{equation}
    |H(\omega)|^2 = \frac{1}{1 + \left( \frac{\omega}{\omega_c} \right)^{ 2n}} 
    \end{equation}
where $\omega_{c}$ is the cutoff frequency and $n$ is the order of the filter. For the majority of our tests, $\omega_{c}$ was set to 0.073\,\AA$^{-1}$, although we did relax the filter for some experiments, increasing $\omega_{c}$ up to 0.12\,\AA$^{-1}$ to ensure that comparisons were not simply products of over-filtering. Although increasing $n$ sharpens the filter, we held it at 5 to avoid significant ringing.
 
While alternative techniques such as the Savitzky-Golay filter are known for their ability to preserve sharp features \citep[see][]{savitsky}, we found that achieving adequate emission/absorption line preservation often required high polynomial orders, which in turn limited the denoising capability of the filter. In contrast, the Butterworth filter provided a better trade-off between smoothing and signal fidelity, with intuitive control over attenuation in the frequency-domain via $\omega_c$. We adopt the {\sc SciPy} \citep[see][]{scipy} implementation of the Butterworth filter as a classical baseline to better contextualise our deep methods. 

\begin{figure}
\includegraphics[width=85mm]{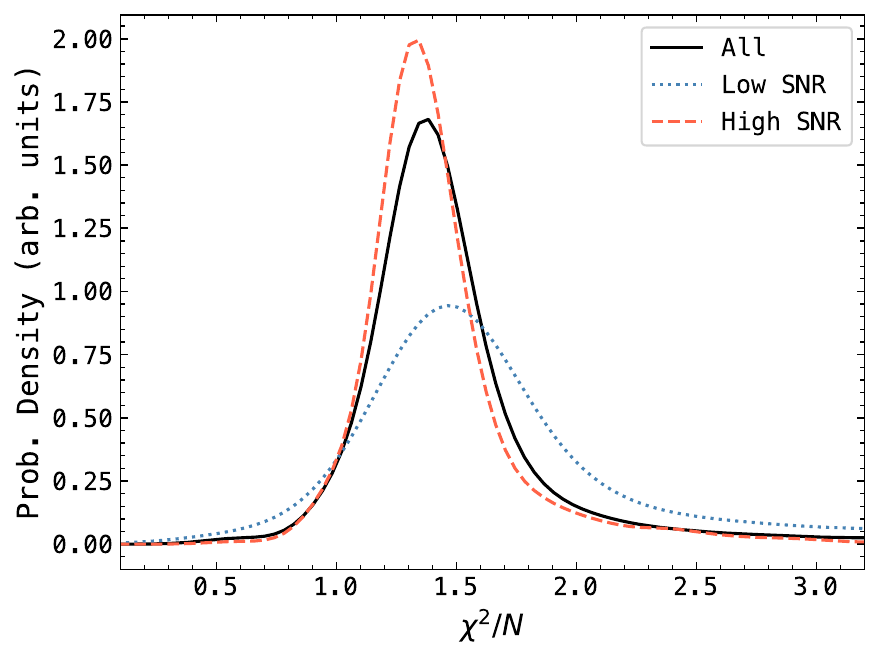}
\caption{Distribution of the $\chi^2$ statistic divided by the number of spectral bins, when comparing the original spectra and the best fit models adopted here as ground truth (see text for details).}
\label{fig:chi2}
\end{figure}

\section{Comparisons of spectral denoising}
\label{Sec:Comp}

A set of 50,000 spectra 
are used to train and validate the models described in the previous section and the recovered (denoised) data are compared with a set of 8,000--10,000 spectra not seen by the models. Once the system is defined, we consider two cases to explore the reconstruction: 1) actual SDSS spectra from the same dataset (i.e. real galaxy spectra); and 2) best fit spectra of unseen SDSS data (i.e. ``noiseless'' model fits) with added noise. Comparing the recovery in both of these cases allows us to assess whether any of the adopted models overfit the training data. 

We emphasize here that the data based on model fits are not used for the training to ensure the denoising procedure does not add unwanted systematics. Note that the fitted data is produced from fits to the original SDSS spectra, thus providing an ideal case where the same ensemble of galaxies is used. The more traditional approach using a grid of models can fail in this respect as the parameter space explored need not map the true sample. The best fit spectra of case 2 are taken from the official spSpec data model from SDSS-Legacy \citep[see, e.g.,][]{DR8}. These correspond to the official SDSS pipeline best fit models for redshift and classification purposes. These best fits are obtained from fits to templates selected from the actual data using Principal Component Analysis \citep[see, e.g.][]{Bolton:12}. The accuracy of best-fit models in reproducing observed galaxy spectra has been well established in previous studies \citep[e.g.][]{CS:99,Yip:04}, and forms an integral part of the Sloan Digital Sky Survey data processing pipeline. Although, as discussed in the introduction, autoencoders and other machine learning methods can reconstruct spectra using a smaller number of components and potentially more flexible representations, PCA remains a highly accurate and robust method for describing observed spectra. Visual inspection confirms these fits provide a good representation of the actual data, and we note that for our purposes we do not need to relate these fits to specific parameters such as age, metallicity, etc. Fig.~\ref{fig:SDSSFits} shows a few of the best-fit models (black lines) along with the original data (grey lines), with the labels identifying the SDSS spectrum (with plate, modified Julian date and fibre ID), along with the median value of the S/N in the SDSS-$g$ band. For reference, we also include the spectral fitting result in each case, using the code {\sc pPXF} \citep{ppxf}. The fitting adopts the E-MILES stellar population synthesis models with the standard range of parameters \citep{E-MILES}. While the official SDSS reconstruction is based on a careful PCA reconstruction \citep{CS:99,Yip:04}, for our purposes, the results are comparable to a more detailed (but model independent) fitting. Moreover, we show in Fig.~\ref{fig:chi2} the distribution of the $\chi^2$ statistic between observational data and SDSS-official best fit model, divided by the number of spectral bins, for a comparison in the full spectral range. The histogram is shown for a general sample (black) and for segregated subsets at the 25\% and 75\% percentile level regarding S/N (as measured in the SDSS-$r$ band). We confirm that these fits are acceptable to be considered, for our purposes, as ground truth in the analysis. However, we note the above mentioned point regarding the use of PCA as a way to obtain these fitted models.

In order to quantify the performance of the models, we opt for a figure of merit that targets the residual of three key line strengths. While a full spectral residual is a valid description, our approach focuses on the use of DL methods in optimising the analysis of stellar population parameters. Previous work on real and synthetic galaxy spectra reveals a rather small set of spectral windows where information is encoded \citep{Entropy}. These include the “blue” and “red” regions defined earlier. Unsurprisingly, these windows happen to be the intervals where traditional line strengths were defined. In this paper we focus on three of the most prominent line strengths: the 4000\AA\ break \citep[as defined by][]{Balogh:99}, the fine (i.e. narrow) definition of H$\delta$ Balmer absorption of \citet{WO:97}, and the traditional Mgb index of the Lick system \citep{SCT:98}.

The figure of merit ($\Delta$) is defined for each line strength measurement as follows: we produce a vector with the residuals of the output and the reference for each spectrum: 
$\delta_{s,r}({\cal I}_i)\equiv[{\cal I}_{i,s}-{\cal I}_{i,r}]$, where $s$ represents the output spectrum and $r$ the reference spectrum. We note that in all our comparisons, the mean/median values of 
$\{\delta_{s,r} ({\cal I})\}$ are always close to zero, as expected. However, the scatter about the mean/median is substantially higher, i.e. we are dominated by this scatter.
The standard deviation gives instead an adequate representation of how well the data are recovered. Therefore, we adopt the standard deviation of the residuals as our figure of merit:
\begin{equation}
\Delta_r ({\cal I})\equiv\sqrt{\langle\delta_{s,r}^2({\cal I})\rangle_s-\langle\delta_{r,s}({\cal I})\rangle^2_s}.   
\end{equation}
This parameter is estimated over the ensemble of all spectra ($\langle\cdots\rangle_s$).
Also note that the reference case (denoted by the subscript $r$) is the comparison spectra, that can be defined in two ways, it is either the (``noiseless'') best fit data ($\Delta_O$), or the noisy original SDSS data ($\Delta_N$). The former allows us to quantify the denoising process, whereas the latter is used to test overfitting. 

\begin{figure}
\includegraphics[width=80mm]{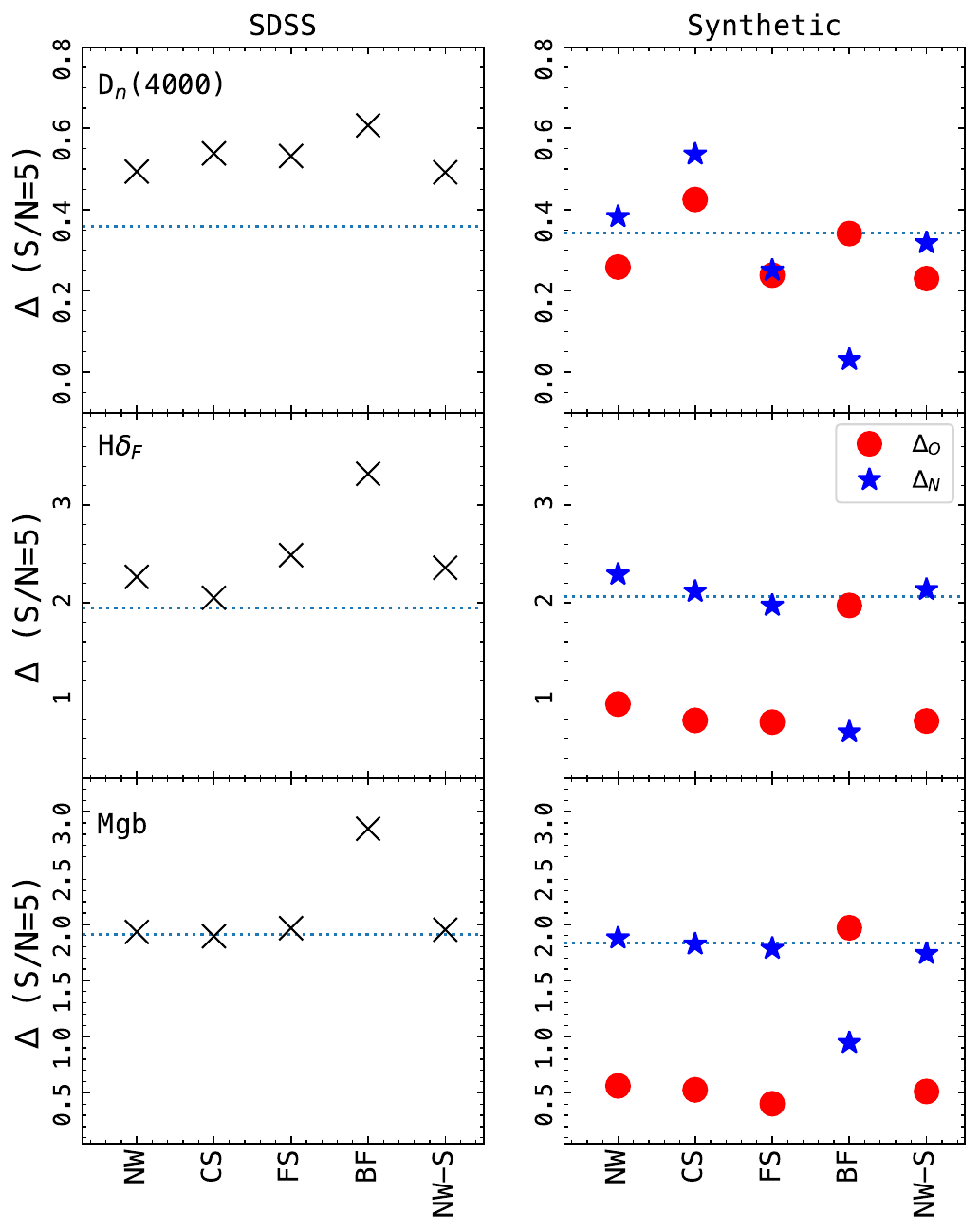}
\caption{Residual statistic ($\Delta$) estimated at S/N=5 for the reconstruction of SDSS galaxy spectra (left) and fitted data (right). The horizontal dashed line represents the residual $\Delta$ for the comparison between the observed data and the best fit spectra. See text for details.}
\label{fig:SN5}
\end{figure}

\begin{figure*}
\includegraphics[width=80mm]{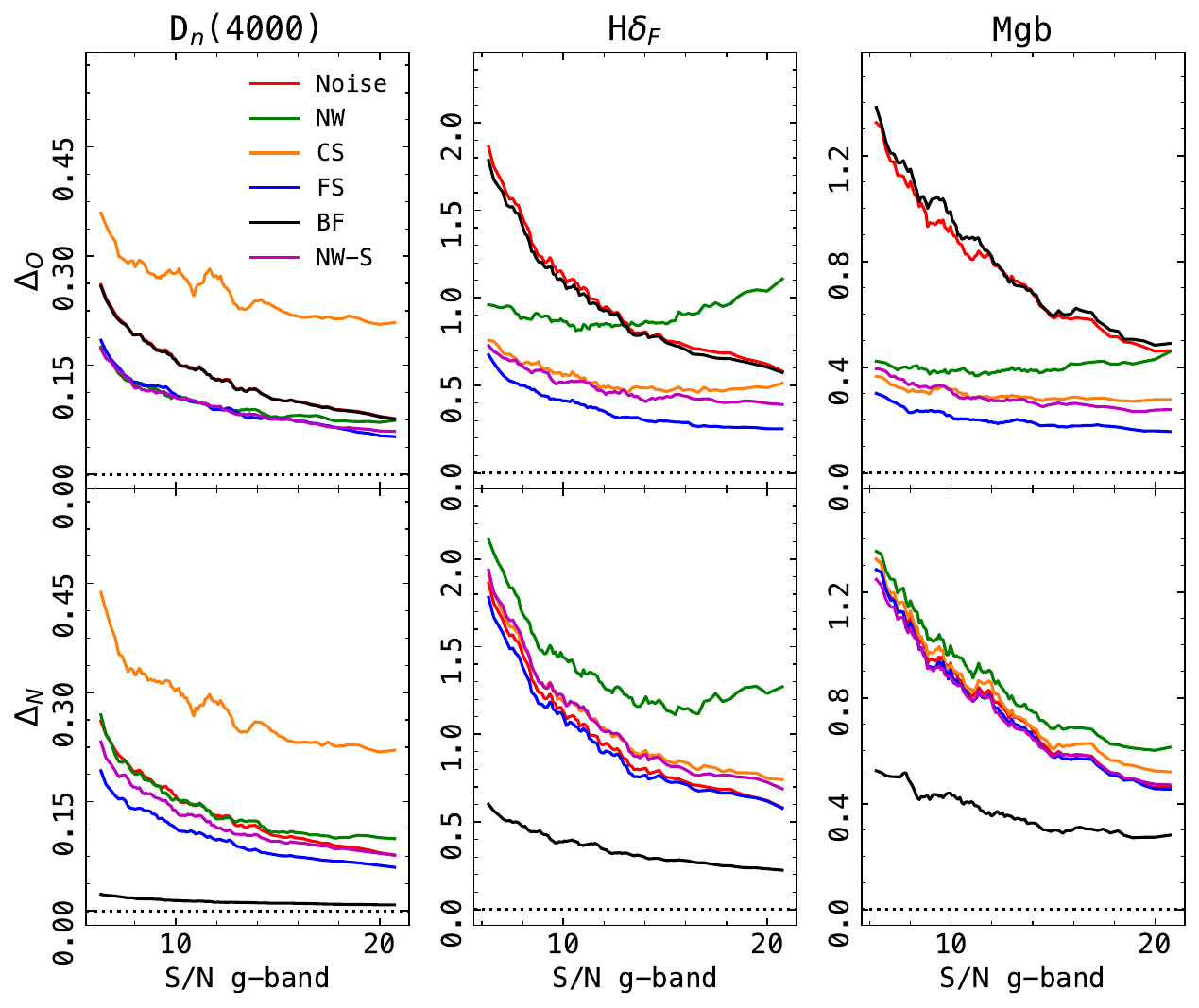}
\includegraphics[width=80mm]{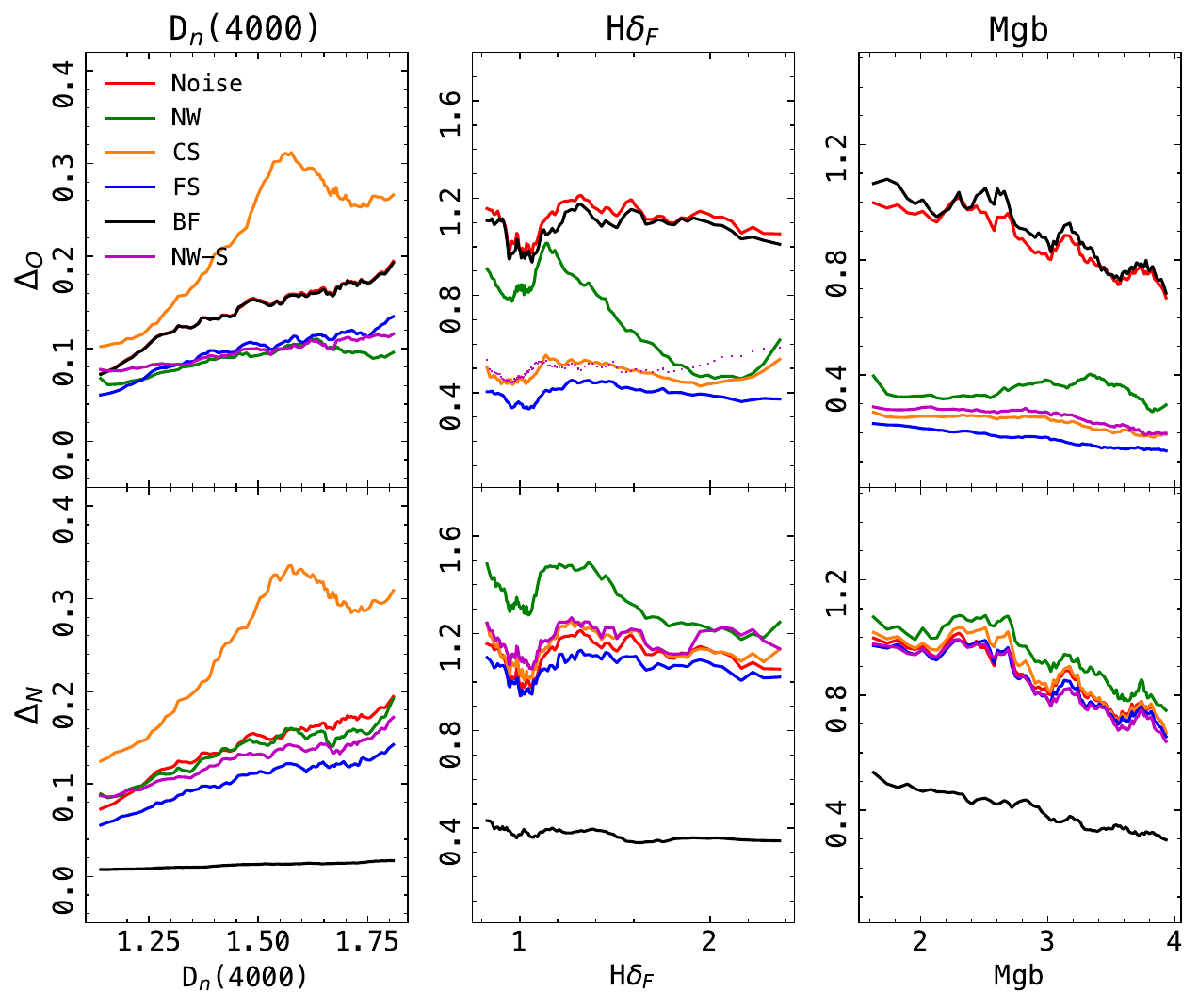}
\caption{Standard deviation of the residuals of three line strengths, 
as labelled, showed with respect to the S/N in the SDSS-$g$ band (left) and the actual line
measurement (right). They correspond to fitted data with added noise (see text for details). The comparisons are made between the recovered spectra and the
original, noiseless data ($\Delta_O$, top) or the noisy input ($\Delta_N$, bottom).}
\label{fig:resid}
\end{figure*}

Fig.~\ref{fig:SN5} shows the residual statistic measured at a S/N=5 (as an average over all spectra with S/N$\in$[5,6], measured as an average in the SDSS-$g$ band) for the reconstruction of an additional, unseen set of real SDSS spectra (left) or a set of synthetic spectra (right). In the case with real data, we only show the data points corresponding to $\Delta_O$, as $\Delta_N$ would trivially compare the noisy data with itself. For the synthetic case, we can compare the residual statistic for the ``ground truth'' case ($\Delta_O$), and for a noisy realization of this ideal case, with Gaussian noise that shows the same S/N as the original data ($\Delta_N$)\footnote{The noise added in each flux bin corresponds to the same S/N per bin of the original spectra, so that the noise model is consistent with the real data.}. In this framework, the case $\Delta_N<\Delta_O$ would be indicative of a kind of overfitting. The opposite would be suggestive of true denoising. For reference, the value of $\Delta_O$ for the comparison of noiseless and fitted data plus noise -- i.e. the variance expected by the presence of noise in the spectra -- is shown in each case as a horizontal dashed blue line. The performance of the different methods is comparable with the SDSS data reconstruction (left panels), although the BF method appears to fare worse.

The more interesting results are found for the fitted data (right panels), where we can discriminate between the recovery of the input, noisy data ($\Delta_N$, blue stars), or a more desirable reconstruction of the original, noiseless, spectra. ($\Delta_O$, red circles). One thing that stands out quite clearly is that the 4000\AA\ break strength is poorly determined by the CS method. This may be quite expected, as the D$_n$(4000) is wider than the other two and relies on the continuum. However, we performed this test as there is a well-known degeneracy between parameters, so that, for instance,  D$_n$(4000) indices are correlated with, e.g, Mgb. The strong covariance between line strengths found in \citet{Entropy} indicates that the absorption line spectrum would encode similar information as the continuum. The large covariance, and the relatively small number of data points (spectral bins), results in an effectively lower number of degrees of freedom,  posing a challenge for DL methods when exploring galaxy spectra.

\begin{figure}
\includegraphics[width=80mm]{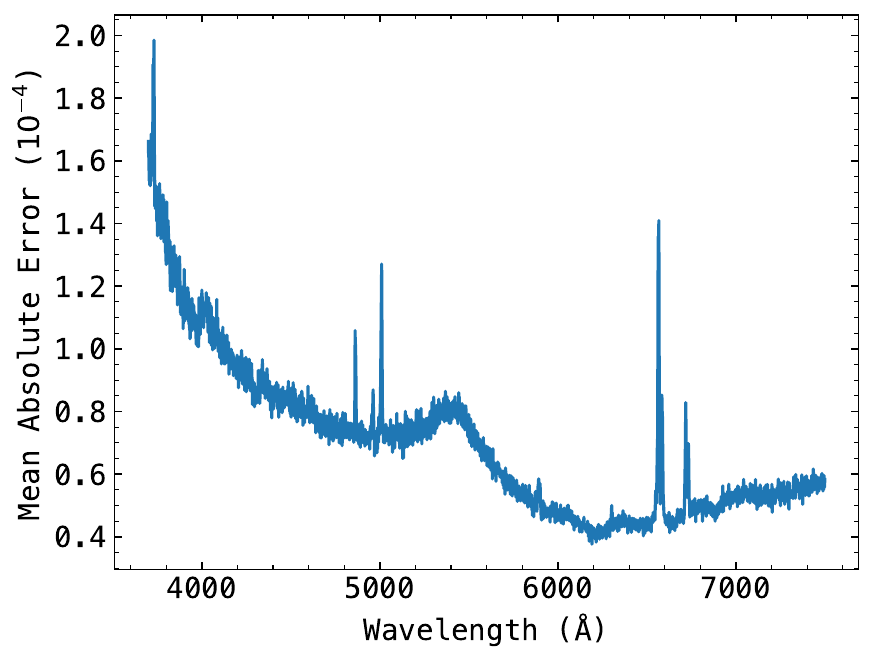}
\caption{The limitations of the CS model are illustrated using the mean errors for each wavelength. To produce this plot, the absolute differences between model reconstructions and corresponding ground truth SDSS test set spectra were computed and then averaged.}
\label{fig:cont_error}
\end{figure}

The figure also shows an uncanny inversion of the star - circle order (i.e. $\Delta_N$ vs $\Delta_O$) in the BF method with respect to the other algorithms. We emphasize that this method does not use information from the ensemble, and only relies on a careful filtering of high frequencies, many of which would be ascribed to noise. This would be equivalent to a truncation in a Fourier series. The results presented here reveal that the BF method tends to overfit, so that it produces an optimal reconstruction when using noisy data as input, but underperforms when the noiseless fitted data are considered. Despite their differences, deviations with respect to the horizontal line indicate that all deep methods achieve some degree of denoising capability, demonstrating the generality of denoising through information bottlenecks. We would also like to note that these models all achieved relatively low average reconstruction losses, with both test and validation on the order of $10^{-5}$. Nevertheless, this examination of key strengths makes it clear that the FS model is favoured – a finding that we will explore in more detail within the next figure.

\begin{figure}
\includegraphics[width=80mm]{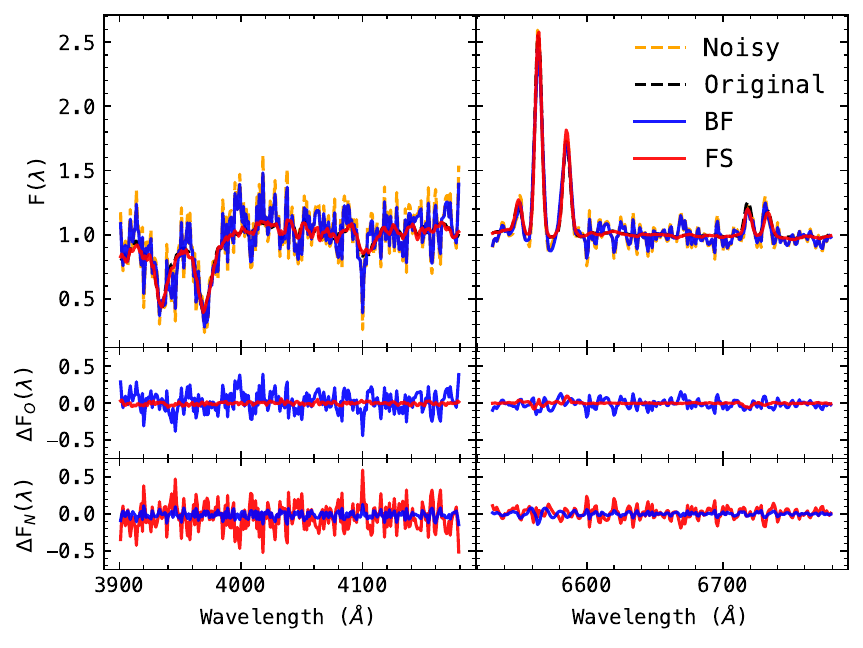}
\caption{Example of the recovery of a spectrum with original S/N$\sim$5. Note how the Butterworth Filter (BF) minimises the residuals with respect to the input (noisy) data, i.e. overfits, whereas FS improves the residuals with respect to the original (noiseless) spectra.}
\label{fig:spec}
\end{figure}

The results for the reconstruction of the full batch of 10,000 noisy fitted data is shown in Fig.~\ref{fig:resid} as a running median when sorting the test sample with respect to the overall S/N (measured in the SDSS-$g$ band, left) or with respect to the measurement of the respective line strength (right). The latter gives an indication of the performance of the denoising with respect to the type of galaxy: for instance, younger stellar populations roughly correspond to a lower value of the 4000\AA\ break strength. Please note that in order to focus more on the bluer part of the spectrum, where the signal tends to be weaker in most galaxies, we opt for the S/N averaged in the SDSS-$g$ band (instead of the $r$ band adopted in the threshold imposed for the sample, as discussed in \S~\ref{Sec:Sample}).

Both $\Delta_N$ (bottom panels) and $\Delta_O$ (top panels) are shown, for the same three line strengths. The characteristic decrease of the residual statistic is evident with respect to increasing S/N. The strong trend of $\Delta$ with D$_n$(4000) is also expected, as stronger breaks imply fainter flux in the blue at around $\lambda\sim$3,800\AA\ where the signal is weaker. The overfitting of the BF method (black lines) produces the lowest $\Delta_N$ (lower panels), but a residual statistic $\Delta_O$ that is comparable with the noisy data, i.e. the Butterworth-reconstructed spectra closely resemble the input noisy SDSS data, but not the original, noiseless data. Out of the other deep methods that appeared to perform similarly in Fig.~\ref{fig:SN5}, the FS (blue) appears to be optimal. The subpar performance of the CS model is also clear as in the previous plot, and so this model is discarded going forward. To better understand where it fails, Fig.~\ref{fig:cont_error} is included as a visualisation of the average error as a function of wavelength. Emission lines and blue wavelengths appear particularly troublesome. This is expected, especially at the blue end of the spectrum where the dominant feature, D$_n$(4000), is associated with particularly high variance. This figure will be explored further in Section~\ref{Sec:explainability}.

Fig.~\ref{fig:resid} reveals that the NW and NW-S exhibit inferior denoising, producing higher residuals than FS in most cases. They are, however, attempting to fulfil a significantly more challenging reconstruction objective. Additionally, their bottleneck differs in that it does not arise from an autoencoder-style constriction, but a lack of information inherent to the training data. The skip connection within NW-S appears to assist in utilising this limited information; while it is interesting to see skip connections providing a small benefit in this context, the limited advantage of NW-S highlights what is potentially a fundamental difficulty in this kind of recovery. As in the case of CS, the NW architectures are excluded from further analysis.

As an illustration of the difference in the reconstruction, we show in Fig.~\ref{fig:spec} details of a noisy spectrum (S/N$\sim$5) in two characteristic wavelength intervals, around the 4000\AA\ break (left) and in the red region where prominent emission lines are present: the H$\alpha$ and [N{\sc II}] complex and the [S{\sc II}] doublet. 
We show the noiseless best fits, the noisy fitted data, and two of the methods: BF and FS. To ease the comparison, in addition to the standard representation with flux (top), we include the residual with respect to the noiseless spectrum (middle panels) and the noisy one (lower panels). As anticipated, BF produces smaller $\Delta F_N$ residuals, but higher ones in $\Delta F_O$. The FS method shows an ideal behaviour and appears to be able to use information from the ensemble to minimise the differences with respect to the ground truth. Using the alternative, gentler set of BF parameters defined in Section~\ref{Sec:Method}, the fundamental issue of over-fitting to $\Delta F_N$ remained.

\begin{figure}
\includegraphics[width=80mm]{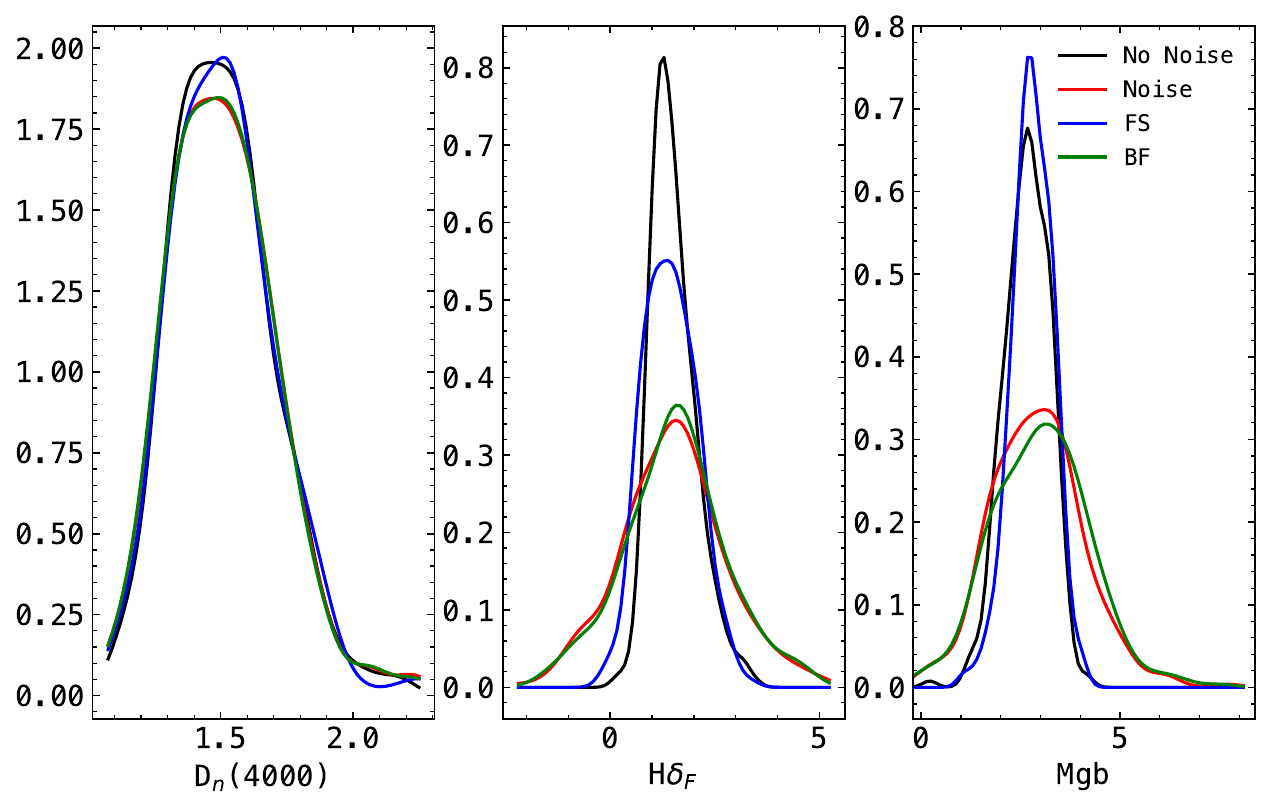}
\caption{Distribution of the three targeted line strengths, for the original SDSS spectra, the noiseless fits and two of the methods explored, as labelled.}
\label{fig:hists}
\end{figure}

A final test to assess a possible systematic in the reconstruction is a comparison of the actual distribution of line strengths in the noiseless and noisy cases as well as the reconstructions. This is shown in Fig.~\ref{fig:hists}, where the histograms of the three line strengths are shown for the same cases as in the previous figure. The small difference in the distribution of D$_n$(4000) is a consequence of the robustness of this index with respect to S/N. Indeed, this is one of the better indices to consider even if the S/N is not high, as the width of the index and the large contrast with respect to the stellar population parameters ensures a meaningful estimate even at S/N$\sim$5. The other two indices show a starker contrast in behaviour between BF and the DL-based method. Note that both H$\delta_F$ and Mgb produce distributions closer to the ground truth with FS, whereas BF closely resembles the wider shape of the histogram of noisy data. From these tests, we conclude that a DL approach is successful at improving the quality of galaxy spectra if trained on a representative, large ensemble with a wide range of S/N, including high quality data with the same instrumental / data characteristics. 


\section{Explaining Model Behaviour}
\label{Sec:explainability}
\begin{figure}
\includegraphics[width=85mm]{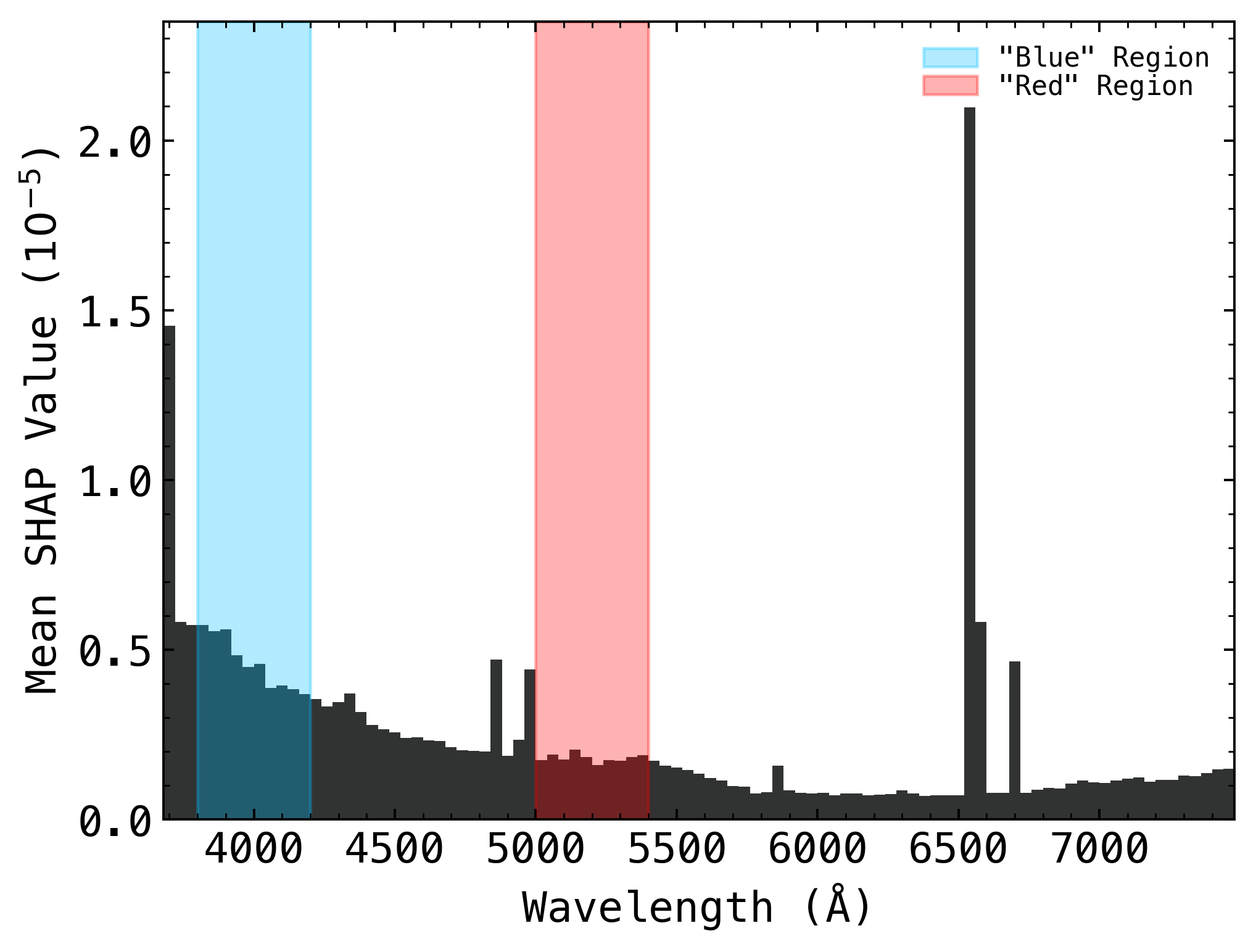}
\caption{Using the FS model, mean SHAP values for flux are plotted as a function of binned wavelength (bin size = 40). The SHAP scores contained within the ``blue'' and ``red'' wavelength windows of interest are highlighted.}
\label{fig:SHAP}
\end{figure}

Deep models are often considered ''black boxes'' as their mappings and decision-making are notoriously difficult to interpret. To better understand how spectral features are being leveraged, we employ SHAP \citep[SHapley Additive exPlanations,][]{shap}. SHAP is a game-theoretic method that explains model outputs by assigning each feature an importance value, based on its contribution to the gap between the model’s actual prediction and its mean prediction, averaged over all possible feature subsets. Fig. \ref{fig:SHAP} shows the mean SHAP scores corresponding to input flux for the trained FS model. In this case, the ''prediction'' is the compressed, latent space encoding. While emission lines like H$\alpha$ and [N{\sc II}] clearly dominate as individual features, it is interesting to note that the importance and therefore predictive power of the continuum is biased considerably towards bluer wavelengths.

These findings are in broad agreement with the negentropy-based analysis of \citet{Entropy}, although the observed variation in SHAP scores suggests that ''useful'' information is much more broadly distributed across the continuum. SHAP has its limitations, and care must be taken when interpreting scores. Despite this, the observed discrepancy highlights the value of considering alternative proxies for information content beyond variance. Entropy-based techniques and classical data-driven methods such as PCA scale the variance in a way that may overlook subtle yet important dependencies within spectral data. Given the SHAP dominance of emission lines, such subtleties may be encouraged by masking these features in future studies and therefore forcing models to leverage more obscure patterns.

The figure also highlights the scores associated with the red and blue regions. While previous research has suggested the strong information content of the former in terms of variance, it does not contain a particularly noteworthy share of the SHAP importance; the Fe and Mg features within this window appear to play a surprisingly minimal role. Nonetheless, the plot makes it clear that their combined input still acts as a useful signal to the FS model, providing insight into how the NW models achieve some ability to predict some unseen spectral regions, as would be expected by the high level of ``information'' entanglement in the absorption spectra across a wide range of wavelengths.

Finally, an important consideration is the overall shape of the plot, which appears to closely mirror the error distribution shown in Fig.~\ref{fig:cont_error}. In other words, the wavelengths that the CS model fails to represent accurately tend to hold strong predictive power in the FS model. This lucidly illustrates the importance of the continuum as a carrier of crucial information - information that appears to pose a surprising challenge for deep reconstruction methods, despite established correlations between the continuum and absorption/emission lines.

\section{Conclusions}
\label{Sec:Conclusions}

\begin{figure}
\includegraphics[width=80mm]{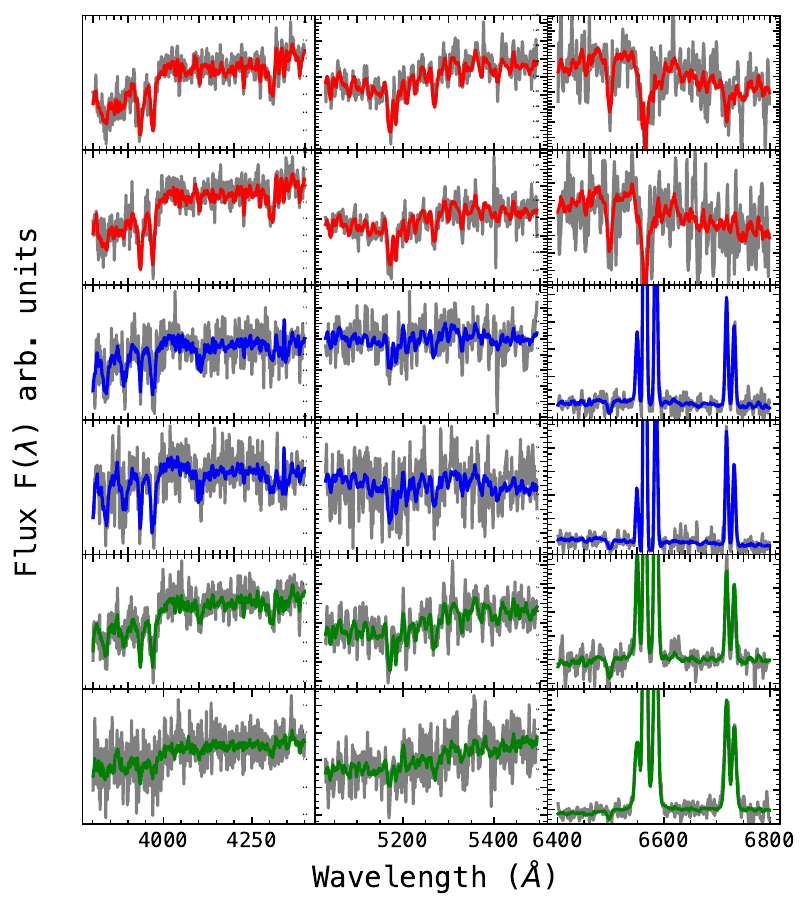}
\caption{Illustration of the recovery with the FS architecture, in three characteristic spectral windows, as labelled. The shaded regions represent real galaxies at low S/N from SDSS and the lines show the recovery. A few cases are shown for galaxies classified as quiescent (red), star-forming (blue) or AGN (green), following the same criteria as in \citet{ZS:23}, based on the classification of the nebular emission lines.}
\label{fig:Somespec}
\end{figure}

This study highlights both opportunities and challenges when applying DL methods to the denoising and reconstruction of galaxy spectra. A key insight is that reproducing spectra through some form of information bottleneck with good generalisation limits the capacity to recreate random noise. Additionally, despite a lack of use within related literature, an MAE reconstruction loss appears to enhance the quality of denoised spectra. Despite the challenging recovery tasks faced by the CS and NW models, all proposed deep methods achieved low average training and validation losses with some ability to denoise. The FS model performed strongest, demonstrating that a standard autoencoder can serve as a practical tool to increase the S/N of both an SDSS training set as well as unseen spectra. This effect can be seen in Fig.~\ref{fig:Somespec} which shows a comparison of the recovered (in colours) and the original (grey) noisy SDSS spectra of six typical cases. The colour coding corresponds to their evolutionary classification as quiescent (red), star forming (blue) and AGN (Green), following the classification of \citet{ZS:23} based on the nebular emission properties. The figure zooms in three interesting spectral regions commonly used for the analysis of the stellar and gaseous phase of galaxies. 

\begin{figure}
\includegraphics[width=80mm]{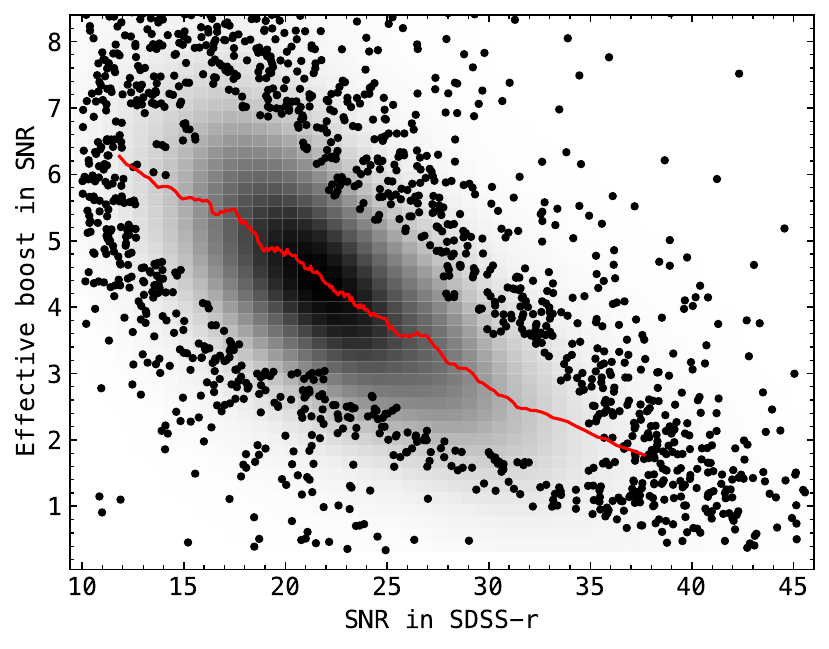}
\caption{Effective increase of S/N in the denoising process, shown as a function of the original S/N (averaged in the SDSS-$r$ band). The red line traces a running median. The data points are shown as a greyscale density plot except for the regions where the number density of points is low. See text for details.}
\label{fig:SNRInc}
\end{figure}

Fig.~\ref{fig:SNRInc} shows a simple assessment of the effective increase in S/N using our methodology. The S/N ``effective boost'' is defined as the ratio between the standard deviation of the input (observed) spectra and that of the ground-truth data -- the latter produced by the intrinsic variations caused by the absorption lines. This estimate is restricted to the 5000-5500\AA\ region, where the continuum is relatively flat and the absorption lines of the stellar populations are prominent, so that this ratio can be effectively interpreted as an increase in S/N. The original S/N (averaged in the SDSS-$r$ band) is plotted in the horizontal axis. The red line traces a running median. While the scatter is substantial, we can conclude that our methodology increases, on average, the S/N of spectra by a factor of $\sim$5 in the noisier data. Needless to say, these results should be considered as ensemble averages, and the reader should be warned against taking these methods at face value on {\sl individual} spectra. We note that the S/N training data was good enough (S/N$\gtrsim$10) to produce meaningful results, and also comprised a wide range in S/N (see horizontal axis of Fig.~\ref{fig:SNRInc}) in order to ``connect'' the low- and high-S/N regime. A more extensive analysis (Camilleri et al, in preparation) is planned with more data, covering a wider range in S/N using the spectra from DESI \citep{DESIDR1}. The astrophysical implications mainly concern more accurate estimates of parameters such as stellar velocity dispersion, or stellar population properties, where the S/N is expected to be high \citep[see, e.g.][]{FLB:13,Woo:24}.

Compared to traditional denoising algorithms, which require careful parameter tuning to balance smoothing with the preservation of sharp features, our approach leverages statistics learned from a vast catalogue of spectra. This minimises biases, limited, of course, to the parent sample. Experiments using fitted data yield quantitative demonstrations that classical filtering techniques are susceptible to smoothed outputs which do not always represent the underlying spectrum. Careful examination of key spectral regions suggest that our DL framework is stronger at recovering this true signal, provided that the noisy input is processed to be consistent with training data. Furthermore, it stands distinct from other deep denoising models: unlike supervised approaches that rely on synthetic spectra with added Gaussian noise, our autoencoder achieves effective noise suppression using only real spectral data without an explicit denoising objective or complex adaptations.
\\
Beyond denoising, our SHAP analysis provides insight into how the models utilise different spectral features, highlighting differences from traditional information-theoretic techniques which target variation, whether or not it is ``useful''. The dominance of specific regions like H$\alpha$ suggests a potential route for further investigation: by selectively reducing the influence of such features, it may be possible to encourage deep models to explore more subtle correlations or patterns. More generally, this line of investigation illustrates the purpose and utility of employing explainability methods within astronomy. As DL is increasingly adopted within the field, it is crucial to maintain the capacity for human understanding.
\\
Looking ahead, several avenues for improvement remain. The standard MAE loss treats all wavelengths as equally important. Instead, biasing the loss to focus on physically important regions could result in better performance. In addition, neural networks are known to be biased towards learning low-frequency signals, as shown by \cite{bias}. Strategies, such as multi-stage neural networks, may assist in representing and utilising sharp features like emission lines. Finally, the adoption of more sophisticated DL approaches, like attention mechanisms \citep[see][]{attention}, could better leverage the subtle correlations found within galaxy spectra.

\section*{Acknowledgements}
OC would like to thank Dr Emily Gray for her useful feedback. ZS and IF acknowledge support from the Spanish Research Agency of the Ministry of Science and Innovation (AEI-MICINN) under grant PID2019-104788GB-I00. Funding for SDSS-III has been provided by the Alfred P. Sloan Foundation, the Participating Institutions, the National Science Foundation, and the U.S. Department of Energy Office of Science. The SDSS-III web site is \href{http://www.sdss3.org/}{http://www.sdss3.org/}.

\section*{Data availability}
This work has been fully based on publicly available data: galaxy spectra were retrieved from the SDSS \href{https://www.sdss.org/dr17/}{DR17 archive} and stellar population synthesis models can be obtained from the respective authors.

\bibliographystyle{mnras}
\bibliography{Denoise}

\appendix

\section{A note on the effect of the loss function}
\label{app:appendixA}

\begin{figure}
\includegraphics[width=80mm]{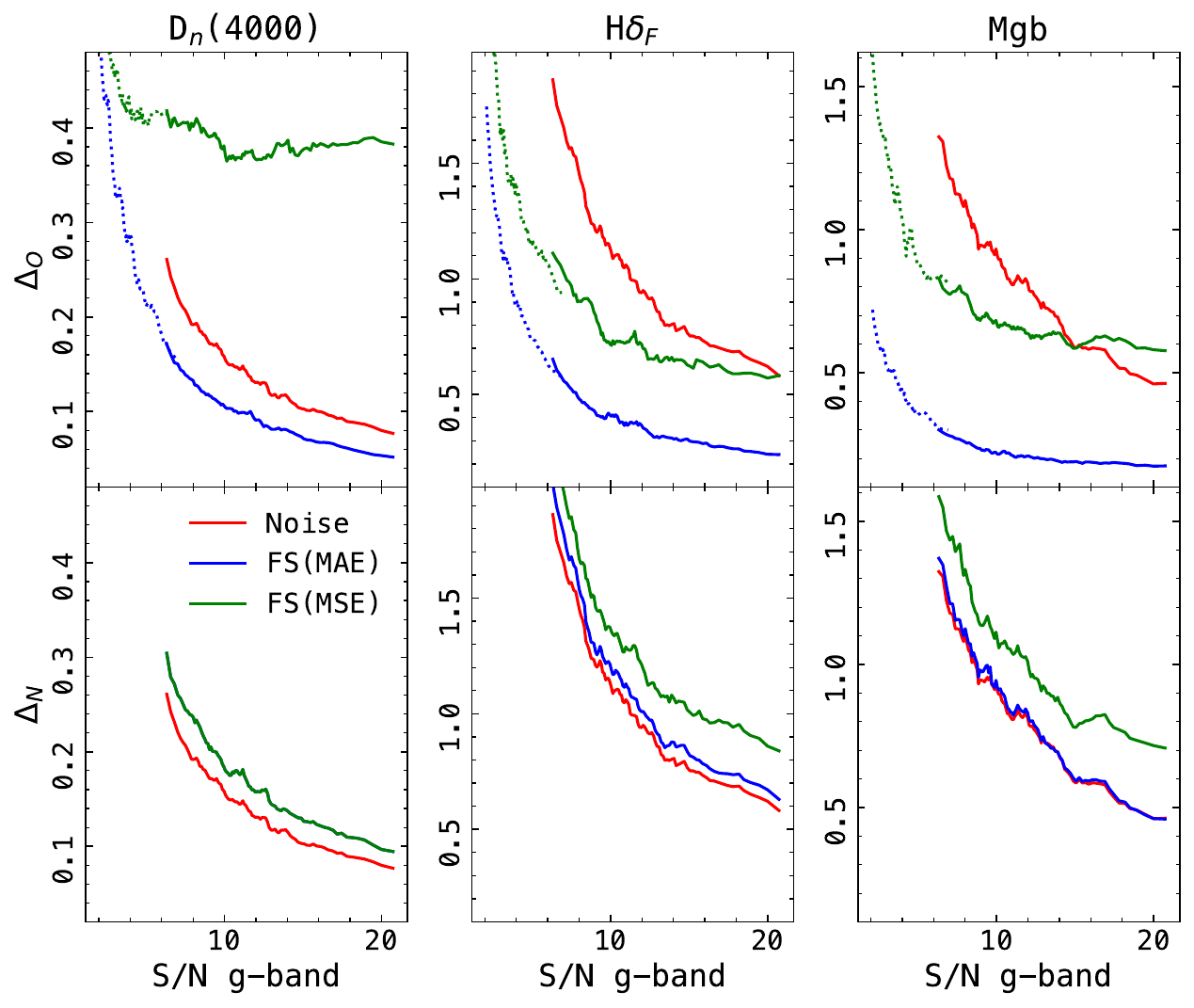}
\caption{Comparison of performance with two models that have the same
architecture (FS) but different loss function, as labelled. The fitted data in this 
case has a noise model corresponding to a Laplacian distribution.}
\label{fig:lapl}
\end{figure}

Among the many details in a DL model is the choice of the loss function. In this paper, we show that a standard architecture with a MAE loss function optimally performs in producing spectra at a higher S/N. However, the comparisons were made with fitted data with noise being simplified to a Gaussian distribution with mean and standard deviation given by the observed flux and uncertainty in each spectral bin. Here we explore an additional case where   
the noise is modelled by a Laplacian distribution with scale given by the standard rule $\lambda=\sigma/\sqrt{2}$. Fig.~\ref{fig:lapl} compares the results when Laplacian noise is added. We also show the same procedure for an identical DL architecture (FS), where the loss function is changed to MSE, which has a higher dependence on outliers. Furthermore, we included an additional batch of noisy data, where the original S/N was decreased by a factor of 3, presented as a dotted extension in the top panels. Even in the case of Laplacian noise (representing a higher number of ``outliers'' with respect to a Gaussian distribution), the FS(MAE) method performs well, whereas FS(MSE) produces noticeably higher residual statistic. We note that the result is equally favourable for MAE even in the case of Gaussian noise, therefore we favour the use of MAE as a robust loss function.

\bsp	
\label{lastpage}

\end{document}